\documentclass[12pt,english,preprint]{emulateapj}
\usepackage[T1]{fontenc}
\usepackage[latin1]{inputenc}
\usepackage{graphicx}
\usepackage{amssymb}

\providecommand{\tabularnewline}{\\}

\usepackage{array}

\makeatletter

\usepackage{times}

\shorttitle{Dynamical evolution of the S-stars}
\shortauthors{Perets et al.}

\makeatother

\usepackage{babel}

\usepackage{babel}

\begin{document}
\global\long\global\long\def\Ms{M_{*}}
 \global\long\global\long\def\LS{L_{*}}
 \global\long\global\long\def\Rs{R_{*}}
 \global\long\global\long\def\Mo{M_{0}}
 \global\long\global\long\def\Lo{L_{0}}
 \global\long\global\long\def\Ro{R_{0}}
 \global\long\global\long\def\Mbh{M_{\bullet}}
 \global\long\global\long\def\SgrA{Sgr\, A^{\star}}

\global\long\def\np{n_{p}}
 \global\long\def\Np{N_{P}}
 \global\long\def\Mp{M_{p}}
 \global\long\def\Ns{N_{\star}}
 \global\long\def\rMP{r_{\mathrm{MP}}}


\title{Dynamical evolution of the young stars in the Galactic center:\\
 N-body simulations of the S-stars }

\author{Hagai B. Perets\altaffilmark{1}, Alessia Gualandris\altaffilmark{2},
Gabor Kupi\altaffilmark{1}, David Merritt\altaffilmark{2} and
Tal Alexander\altaffilmark{1} }

\altaffiltext{1}{Weizmann Institute of Science, POB 26, Rehovot
76100, Israel} \altaffiltext{2}{Department of Physics and Center
for Computational Relativity and Gravitation, Rochester Institute
of Technology, Rochester, NY 14623}

\email{hagai.perets@weizmann.ac.il}
\begin{abstract}
We use $N$-body simulations to study the evolution of the orbital
eccentricities of stars deposited near ($\lesssim0.05$ pc) the Milky
Way massive black hole (MBH), starting from initial conditions motivated
by two competing models for their origin: formation in a disk followed
by inward migration; and exchange interactions involving a binary
star. The first model predicts modest eccentricities, lower than those
observed in the S-star cluster, while the second model predicts higher
eccentricities than observed. The $N$-body simulations include a
dense cluster of $10M_{\odot}$ stellar black holes (SBHs), expected
to accumulate near the MBH by mass segregation. Perturbations from
the SBHs tend to randomize the stellar orbits, partially erasing the
dynamical signatures of their origin. The eccentricities of the initially
highly eccentric stars evolve, in $20$ Myr (the S-star lifespan),
to a distribution that is consistent at the $\sim\!95\%$ level with
the observed eccentricity distribution. In contrast, the eccentricities
of the initially more circular orbits fail to evolve to the observed
values in 20 Myr, arguing against the disk migration scenario. We
find that $20\%$--$30\%$ of the S-stars are tidally disrupted by
the MBH over their lifetimes, and that the S-stars are not likely
to be ejected as hypervelocity stars outside the central 0.05 pc by
close encounters with stellar black holes. 
\end{abstract}


\section{Introduction}

In recent years, high resolution observations have revealed the existence
of many young OB stars in the Galactic center (GC). Accurate measurement
of the orbital parameters of these stars gives strong evidence for
the existence of a massive black hole (MBH) which dominates the dynamics
in the GC \citep{ghe+98,eis+05,gil+08}. Most of the young stars are
observed in the central 0.5 pc around the MBH. The young star population
in the inner 0.05 pc (the {}``S-stars'') consists exclusively of
B-stars, in an apparently isotropic distribution around the MBH, with
relatively high eccentricities ($0.3\lesssim e\lesssim0.95$; \citealt{gil+08}).
The young stars outside this region comprise O-stars in one or two
disks, and present markedly different orbital properties \citep{lev+03,bar+08,lu+09}.

Since regular star formation in the region near the MBH is inhibited
by tidal forces, many suggestions have been made regarding the origin
of the S-stars. Many of these are probably ruled out by observations
and/or by theoretical arguments \citep[see][for a
  review]{ale05,pau+06}. The various scenarios for the origin of the S-stars predict very
different distributions for their orbits, which in principle could
be constrained by observations. However, it is not clear to what extent
relaxation processes can produce changes in the distribution of orbital
parameters after the stars have been deposited near their current
locations. Here we try to resolve this question. We use $N$-body
simulations to follow the evolution of stellar orbits around the GC
MBH for $20$ Myr, starting from various initial conditions that were
motivated by different models for the origin of the S-stars. However,
we do not discuss here the possibility that an intermediate mass black
hole was involved in the production and/or evolution of the S-stars,
which is discussed in details elsewhere \citep[see ][and references therein]{mer+09}.
We find our N-body results to be consistent with our analytical predictions,
and compare them with current observations. We then discuss the implications
for the validity of the models for the production of the S-stars.
In addition we study the possible ejection of the S-stars outside
the inner $0.05$ pc and the contribution of such ejected stars to
the population of hypervelocity stars (as suggested by \citet{ole+07})
and to the isotropic population of B-stars observed at distances of
up to $0.5$ pc from the MBH (i.e. at distances similar to those of
the young O/WR stellar disks, but outside of these disks at high inclinations).

In \S2 we summarize the different models for the origin of the S-stars
and the predictions that they make for the stellar orbits at the time
when the stars are first deposited near the MBH. \S3 describes the
$N$-body simulations we carried out to follow the S-star orbital
evolution starting from these initial conditions. \S4 and \S5 present
the results of these simulations and discusses the implications for
the origin of the stellar populations of B-stars in the GC and for
the population of hypervelocity stars observed in the Galactic halo.
\S6 sums up.

\section{Models for the S-stars origin}

Many solutions have been suggested for the origin of the S-stars,
but many of these have been effectively excluded (see \citealp{ale05,pau+06}
for a review). Here we focus on two basic models which differ substantially
in their predictions for the initial orbital distribution of the S-stars
and/or the time passed since their arrival/formation at their current
location. These are (1) formation of the S-stars in a stellar disk
close to the MBH, followed by transport through a planetary-migration-like
scenario to their current positions \citep{lev07}; (2) formation
of the S-stars in binaries far from the MBH followed by scattering
onto the MBH by massive perturbers (e.g. giant molecular clouds) and
tidal disruption of the binaries \citep{per+07,per+08}, leaving a
captured star in a tight orbit around the MBH. Binary disruption scenarios
similar to (2) have been proposed, in which the S-stars formed in
a stellar disk (either the currently observed 6 Myr old disk, or an
older, currently not observed disk) and later changed their orbits
due to coherent torques through an instability of eccentric disks
\citep{mad+08}; or through the Kozai mechanism resulting from the
presence of two disks \citep{loc+08b}. The latter alternative is
unlikely since the Kozai mechanism is quenched in the presence of
a massive enough cusp of stars such as exists in the GC \citep{cha+08,mad+08}).
In any case, all the binary-disruption scenarios imply very similar
initial distributions for the captured S-stars, and may differ only
in their relevant timescales.

In the following, we briefly discuss the initial distribution of the
eccentricities and inclinations of the S-stars expected from the different
scenarios for their production. The models are summarized in Table
\ref{t:models}.

\begin{table*}
\caption{\label{t:models}Models for the S-stars origin }

\centering{}\begin{tabular}{llccccl}
\hline 
\#  & Origin  & Initial  & Time  & Model  & Survival  & Refs.$^{c}$\tabularnewline
 &  & Eccentricity  & (Myr)  & Probability$^{a}$  & Fraction$^{b}$  & \tabularnewline
\hline 
1  & Capture following binary disruption due  & High  & $6$  & $0.26$  & $0.8$  & 1\tabularnewline
 & to disk instability in the currently observed disk  & ($0.94\le e\le0.99$)  &  &  &  & \tabularnewline
2  & Capture following binary disruption due  & High  & $20$  & $0.93$  & $0.7$  & 2\tabularnewline
 & to massive perturbers or to disk instability  & ($0.94\le e\le0.99$)  &  &  &  & \tabularnewline
 & in an old non-observed disk  &  &  &  &  & \tabularnewline
3  & Disk formation + planetary like  & Low  & $6$  & $8\times10^{-3}$  & $0.9$  & 3\tabularnewline
 & migration (currently observed disk)  & ($e\le0.5$)  &  &  &  & \tabularnewline
4  & Disk formation + planetary like  & Low  & $20$  & $0.06$  & $0.8$  & 3\tabularnewline
 & migration (possible old disk)  & ($e\le0.5$)  &  &  &  & \tabularnewline
\hline 
\multicolumn{7}{l}{$^{a}${\footnotesize Probability for the samples of the observed
and simulated S-stars}}\tabularnewline
\multicolumn{7}{l}{{\footnotesize to be randomly chosen from the same distribution (see
text).}}\tabularnewline
\multicolumn{7}{l}{$^{b}${\footnotesize Fraction of S-stars not disrupted by the MBH
during the simulation (see text).}}\tabularnewline
\multicolumn{7}{l}{$^{c}$ (1) \citet{mad+08} (2) \citet{per+07} (3)\citet{lev07}}\tabularnewline
\hline
\end{tabular}
\end{table*}

\subsection{Binary disruption by a massive black hole}

\label{sec:Binary-disruption}

A close pass of a binary star near a massive black hole results in
an exchange interaction, in which one star is ejected at high velocity,
while its companion is captured by the MBH and left on a bound orbit.
Such interaction occurs because of the tidal forces exerted by the
MBH on the binary components. Typically, a binary (with mass, $M_{bin}$,
and semi-major axis, $a_{bin}$), is disrupted when it crosses the
tidal radius of the MBH (of mass $\Mbh$), given by $r_{t}=a_{bin}(\Mbh/M_{bin})^{1/3}$.
One of the binary components is captured by the MBH \citep{gou+03}
on a wide and eccentric orbit while the companion is ejected with
high velocity \citep{hil88}.

The capture probability and the semi-major axis distribution of the
captured stars were estimated by means of numerical simulations, showing
that most binaries approaching the MBH within the tidal radius $r_{t}(a_{bin})$
are disrupted \citep{hil91,hil92,gua+05,bro+06c}. The harmonic mean
semi-major axis for 3-body exchanges with equal mass binaries was
found to be \citep{hil91} \begin{equation}
\left\langle a_{cap}\right\rangle \simeq0.56\left(\frac{\Mbh}{M_{\mathrm{bin}}}\right)^{2/3}a_{bin}\simeq\,0.56\left(\frac{\Mbh}{M_{\mathrm{bin}}}\right)^{1/3}r_{t},\label{e:afinal}\end{equation}
 where $a_{bin}$ is the semi-major axis of the infalling binary and
$a_{cap}$ that of the captured star (the MBH-star {}``binary'').
Most values of $a_{cap}$ fall within a factor $2$ of the mean. This
relation maps the semi-major axis distribution of the infalling binaries
to that of the captured stars: the harder the binaries, the more tightly
bound the captured stars. The periapse of the captured star is at
$r_{t}$, and therefore its eccentricity is very high \citep{hil91,mil+05},
\begin{equation}
e\!=\!1-r_{t}/a_{cap}\!\simeq1-1.8(M_{\mathrm{bin}}/\Mbh)^{1/3}\!\simeq\!0.94-0.99\label{eq:capture-eccentricity}\end{equation}
 for values typical of B-type main sequence binaries and the MBH in
the GC ($M_{bin}=6-30\,\Mo$; $\Mbh=3.6\times10^{6}\,\Mo$; $a_{cap}=0.5-2\times\left\langle a_{cap}\right\rangle $).
Therefore, in order to study the evolution of S-stars from the binary
disruption scenarios we assume that the initial eccentricities of
S-stars are in the range $0.94-0.99$ (where most are close to the
mean value of $0.98$).

In principle the binary disruption scenario has specific predictions
for the semi-major axis distribution of the captured stars, which
could also be used for constraining the model. However such distribution
is highly sensitive to differences in the (unknown) binary distribution
in the GC region. The prediction of high eccentricities for the captured
S-stars, instead, is robust and has only a weak dependence on the
mass of the binary.

The inclinations of the captured S-stars in the massive perturbers
scenario \citep{per+07} are likely to be distributed isotropically
since the stars originate in an isotropic cusp. Although in the disk
instability scenario \citep{mad+08} the progenitors of the captured
stars form in the stellar disk, their original inclinations could
be excited to higher inclinations than the typical inclinations observed
in the stellar disk (Y. Levin, private communication), and may resemble
a more isotropic distribution.

\subsection{Planetary like migration from the young stellar disk(s)}

\citet{lev07} suggested that the S-stars could have formed in the
currently observed stellar disk in the GC \citep{bar+08,lu+09}, and
then migrated inward in a way similar to planetary migration. The
migration time scale expected from such a scenario could be as short
as $10^{5}$ yr (for type I migration), which could be comparable
to (although possibly larger than; \citet{nay+07}) the lifetime of
the gaseous disk. Recent analytic work (\citet{ogi+03,gol+03b}; and
references therein) has shown that eccentricity is likely to be damped
during migration, unless eccentricity excitation occurs, which requires
the opening of a clean gap in the disk. In the latter case the migration
timescale might be larger \citep{lev07}, possibly inconsistent with
the lifetime of the gaseous disk \citep{nay+07}. It is therefore
more likely that the eccentricities of the stars are damped during
the migration. Even eccentricity excitation, if such took place, is
unlikely to excite very high eccentricities. The mean eccentricity
of the observed stars in the stellar disk is $0.34\pm0.06$. Therefore,
in order to study the evolution of the S-stars following their formation
in a stellar disk, we assume them to have low eccentricities, or,
being conservative, moderately high eccentricities ($e_{max}=0.5$;
where we use a thermal distribution of eccentricities, cut off at
$e_{max}$). These simulations include the less likely (\citealt{lev07})
possibility that the stellar disk extended inward to the current region
of the S-stars, in which case the S-stars were formed in-situ and
did not migrate.

\section{The N-body simulations}

To test these competing models, we carried out $N$-body simulations
of the inner Milky Way bulge using models containing a realistic number
of stars. All integrations were carried out on the 32-node GRAPE cluster
\texttt{gravitySimulator} 
 at the Rochester Institute of Technology which adopts a parallel
setup of GRAPE accelerator boards to efficiently compute gravitational
forces. The direct-summation code $\phi$GRAPE was used \citep{har+07}.
The simulations used a softening radius of $\sim4\, R_{\odot}$ comparable
to the radius of the S-stars $r_{\star}$, so as to be able to follow
even the closest encounters between stars.

Our initial conditions were based on a collisionally evolved model
of a cusp of stars and stellar remnants around the GC MBH \citep[hereafter HA06; see also \citealt{fre+06}]{hop+06b}.
HA06 evolved the multi-mass isotropic Fokker-Planck equation representing
the stellar distribution in the region extending to $\sim1$ pc from
the MBH; in their models the contribution to the gravitational potential
from the distributed mass is ignored. HA06 fixed the relative numbers
of objects in each of four mass bins by assuming a mass function consistent
with continuous star formation. The $10M_{\odot}$ stellar-mass black
holes (SBHs) were found to follow a steep, $n(r)\sim r^{-2}$ density
profile near the MBH while the lower-mass populations (main-sequence
stars, white dwarfs, neutron stars) had $n\sim r^{-\alpha}$, $1.4\lesssim\alpha\lesssim1.5$.
The SBHs were found to dominate the mass density inside $\sim0.01$
pc.

Based on this model, we constructed an $N$-body realization containing
a total of 1200 objects within 0.3 pc of the MBH: 200 {}``stars,''
with masses of $3\, M_{\odot}$, and 1000 {}``black holes'' with
masses $10\, M_{\odot}$, around a MBH of $3\times10^{6}\, M_{\odot}$.
We set $\alpha=2$ for the SBHs and $\alpha=1.5$ for the lower-mass
stars, and each density component was tapered smoothly to zero beyond
$0.1$ pc when computing the corresponding $f(E)$. Since the S-stars
may have masses as high as $\sim10\, M_{\odot}$, the higher mass
stars in our simulations could also be treated as S-stars. We did
not see any major differences in the evolution of the more massive
and the less massive stars, and we discuss the evolution of both together.

The number of SBHs contained within a radius $r$ in our $N$-body
models was \begin{equation}
N(<r)\approx600\left(\frac{r}{0.1\ {\rm pc}}\right)\label{eq:nofr}\end{equation}
 implying a distributed mass within $0.1$ pc of $\sim10^{4}M_{\odot}$.
This is somewhat ($\sim2-3\times$) lower than the mass in SBHs in
the HA06 or similar \citep{mor93,mir+00,fre+06} models at the same
radius, and a factor $\sim5$ lower than the \textit{total} mass (mostly
in main-sequence stars) in the Fokker-Planck models. In this sense,
the rates of evolution that we infer below can be considered to be
conservative.

On the other hand, we note that the late-type (old) stars that dominate
the number counts in this region have a much flatter density profile
than predicted by the HA06 models, possibly even exhibiting a central
{}``hole'' \citep{figer+03,zhu+08}. Only the B-type stars in the
nuclear star cluster show a steeply-rising number density, $\alpha=1.1\pm0.3$
\citep{schoe+07,gil+08} but they presumably constitute a negligible
fraction of the total mass in this region, and in any case are far
too young to have reached a collisional steady state around the MBH.
While the origin of this discrepancy between models and observations
is currently unresolved, it may imply that the other contributors
to the distributed mass around the MBH, including the SBHs, also have
a lower density than in the Fokker-Planck models. For instance, relaxation
times at the GC may be too long for collisionally relaxed steady states
to have been established in the last 10 Gyr \citep{mer+06}.

Modeling of the stellar proper motion data \citep{trippe+08,schoe+09}
implies a distributed mass within 1 pc of $0.5-1.5\times10^{6}M_{\odot}$,
but these data are consistent with both rising and falling mass densities
within this region and the distributed mass in the inner $0.1$ pc
is essentially unconstrained \citep{schoe+09}.

Because of these uncertainties, we discuss below how our results would
vary if different numbers of SBHs were assumed.

As discussed in the previous section, we studied two basic sets of
initial conditions for the S-stars. In the first model we assumed
that the S-stars were captured by the MBH as in the binary disruption
scenario \citep{gou+03}, which leaves the captured stars in highly
eccentric orbits ($>0.94-0.99$; cf. section \ref{sec:Binary-disruption}).
Under these assumptions, the stars evolved for $6$ Myr (if formed
in the stellar disk) or longer (if the S-stars formed outside the
central pc, i.e. not in the young stellar disk). We evolved the models
for up to $20$ Myr, which is comparable to the lifetime of the observed
S-stars (although some may have longer lifespans). In the second scenario
we assumed that the S-stars formed in a gaseous disk and then migrated
inwards (or formed in situ in a disk extending close to the MBH).
For this case we assumed the S-stars to have low eccentricities ($<0.5$),
as typical of disk formation models, and to evolve for $6\,$Myr (the
lifetime of the observed stellar disk). In order to check both scenarios
we selected the stars with initially high eccentricity orbits ($0.94\leq e\leq0.99$)
and low eccentricity orbits ($e\leq0.5$) and followed their evolution
for a time appropriate to their presumed origin.

In addition to these evolutionary scenarios we also studied the possibility
of ejection of S-stars as hypervelocity stars, following a close encounter
with a SBH in the vicinity of the MBH, as suggested by \citet{ole+07}.
Our high resolution simulations can accurately follow close encounters
between stars, and therefore track any resulting high velocity ejections
of stars. Motivated by recent observations of B-type main sequence
stars outside the inner $0.05$ pc of the Galaxy, which show evidence
of random orientations similar to those of the S-stars, we also looked
for stars scattered to larger distances by gravitational encounters.
In other words, we considered whether captured S-stars could dynamically
evolve to become the extended B-type stars population observed outside
the central $0.05$ pc.

Although some binaries could exist in the close regions near the MBH
\citep{per08b,per09}, these are likely to be rapidly disrupted and
not play an important role in the dynamical evolution of the S-stars
\citep{per09}. We therefore do not include any binaries in our initial
conditions for the S-stars and SBHs evolution.

\section{Results}

\subsection{Simulations vs. theory: resonant relaxation}

\label{sub:RR}

\begin{figure}[t]

\noindent \begin{centering}
\includegraphics[width=1\columnwidth]{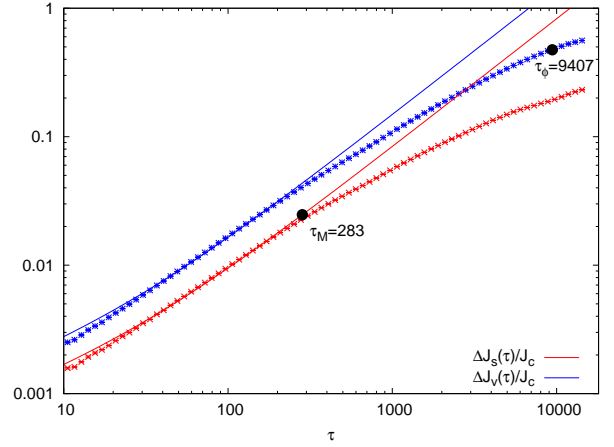} 
\par\end{centering}

\caption{\label{f:RR}The correlation curves $\left|\Delta J\right|/J_{c}$
and $\left|\Delta\mathbf{J}\right|/J_{c}$ as a function of the time
lag $\tau$ for all the stars in the simulation. The simulation data
(points) are compared to the best fit theoretical curves (lines).
Four regimes are seen \citep[see
    text and also]{eil+08}. At very short timescales, non-resonant two-body relaxation ($\propto\!\sqrt{\tau}$)
dominates over RR ($\propto\!\tau$); At $\tau\!<\!\tau_{M}$ the
curves display a coherent, linear rise; At $\tau_{M}\!<\!\tau\!<\!\tau_{\phi}$,
the curves display a random-walk growth and at $\tau\!>\!\tau_{\phi}$
they begin to saturate. }

\end{figure}

We applied the correlation curve method \citet{eil+08} to our simulations
to identify the relaxation process responsible for the dynamical evolution
of the stars (Fig. \ref{f:RR}). The method is able to detect and
measure relaxation in nearly Keplerian $N$-body systems. In the isotropic
system considered here, the angular momentum of the stars, $J$, evolves
both due to the slow stochastic two-body relaxation (e.g. \citealt{bin+87})
and to the rapid resonant relaxation (RR) \citep{rau+96,rau+98,hop+06a,gur+07,eil+08}.
Two-body relaxation changes $J$ in a random walk fashion, $\left|\Delta J\right|/J_{c}\!=\sqrt{\tau/\tau_{NR}}$
over the long two-body relaxation time-scale $\tau_{NR}\!\sim\! Q^{2}/(N\log Q)$,
where $J_{c}$ is the maximal (circular orbit) angular momentum for
a given energy, $Q\!=\!\Mbh/\Ms$, $N$ is the number of enclosed
stars on the distance scale of interest, and time $\tau\!=\! t/P$
is measured in terms of the orbital period on that scale. Resonant
relaxation occurs when the symmetries of the potential act to constrain
the stellar orbits (e.g. closed ellipses in a Kepler potential, or
planar rosettes in a spherical one). As long as the symmetry is approximately
maintained on the coherence timescale $t_{w}$, the stars experience
coherent torques, and $\Delta J/J_{c}\!\sim\!(\sqrt{N}/Q)\tau$. In
a nearly Keplerian potential, as is the case in the inner parsec of
the GC, RR can change both the magnitude of $J$ ({}``scalar RR'')
and its direction ({}``vector RR''). In the Newtonian context, the
coherence of scalar RR is limited by the precession of the apoapse
due to the enclosed stellar mass, on a time-scale $\tau_{M}\!\sim\! Q/N$.
On timescales $\tau\!>\!\tau_{M}$, the coherent change $\Delta J(\tau_{M})$
becomes the mean free path in $J$-space for a rapid random walk,
$\left|\Delta J\right|/J_{c}\!=\!\sqrt{\tau/\tau_{sRR}}$, where $\tau_{sRR}\!\sim\! Q\!\ll\!\tau_{NR}$.
The coherence of vector RR is self-limited by the change in the orbital
orientation due to RR, and is even faster, $\left|\Delta\mathbf{J}\right|/J_{c}\!=\!\sqrt{\tau/\tau_{vRR}}$,
where $\tau_{vRR}\!\sim\! Q/\sqrt{N}$.

Figure (\ref{f:RR}) shows the rms change in the scalar and vector
angular momentum of the stars in the simulation, as a function of
the time lag $\tau$ (correlation curves), up to the maximal time
lag for which the simulation can still be analyzed with high statistical
confidence, $\tau_{\max}\sim10^{4}$. In our simulation%
\footnote{For a spectrum of masses, $\tau_{M}\!\sim\! Q/N\left\langle \Ms\right\rangle $
and $\tau_{sRR}\!\sim\!\Mbh/M_{\mathrm{eff}}$, where $\left\langle \Ms\right\rangle \!=\!8.8\, M_{\odot}$
and $M_{\mathrm{eff}}\!=\!\left\langle \Ms^{2}\right\rangle /\left\langle \Ms\right\rangle $.
$M_{\mathrm{eff}}\!\simeq\!9.7\, M_{\odot}$ for our simulation.%
}, the scalar RR coherence time expected from theory is $\tau_{M}\!\sim\!283$
and the scalar RR timescale is $\tau_{sRR}\!\sim\!3.1\times10^{5}$.
The behavior of the scalar correlation curve clearly indicates that
RR dominates relaxation in our GC model. The turn from the coherent
phase to the random-walk phase of RR is seen at $\tau\!\sim\!300$,
close to the predicted value of $\tau_{M}$. The random-walk growth
continues up to full randomization and saturation at $\sim\!\tau_{\phi}$,
as expected \citep{eil+08}. The scalar RR timescale can be estimated
directly from the simulated data by $\tau_{RR}=\tau/(\left|\Delta J\right|/J_{c})^{2}$;
substituting $\left|\Delta J\right|/J_{c}\!\sim\!0.15$ at $\tau\!\sim\!\tau_{\phi}/2\!\sim\!5000$
yields $\tau_{sRR}\!\sim\!2.2\times10^{5}$, in good agreement with
the predicted value.

Furthermore, since $J/J_{c}\!=\!\sqrt{1-e^{2}}$, the RR-driven eccentricity
evolution relates the final eccentricity, $e_{f}$, after time lag
$\tau$ to the initial eccentricity, $e_{i}$. We therefore expect
the eccentricities at some given time lag to have a typical spread,
$e^{-}(\tau)\lesssim e_{f}(\tau)\lesssim e^{+}(\tau)$, where \begin{equation}
e^{\pm}(\tau)=\left[1-\left(\sqrt{1-e_{i}^{2}}\mp\sqrt{\frac{\tau}{\tau_{RR}}}\right)^{2}\right]^{1/2}.\end{equation}
 The magnitude of the predicted change in eccentricity agrees well
with that observed in the simulations. For example, an S-star initially
captured by a tidal exchange event on a $P=500$ yr ($a\!\sim\!0.04$
pc), $e_{i}=0.97$ orbit can evolve by RR to an $e_{f}\!=\!0.80$
orbit in 20 Myr. The short vector RR timescale $\tau_{vRR}\!\sim\!10^{4}$
($t_{vRR}\!=\!5\times10^{6}$ yr at 0.04 pc) implies full randomization
of the orbital planes after 6 Myr, throughout the S-cluster volume,
as is observed in the simulation. We therefore conclude that RR is
the dominant mechanism responsible for the dynamical evolution of
the S-stars and other stars close to the MBH in the GC.

\subsection{S-star eccentricities and inclinations}

\begin{figure}[!]
\includegraphics[scale=0.3]{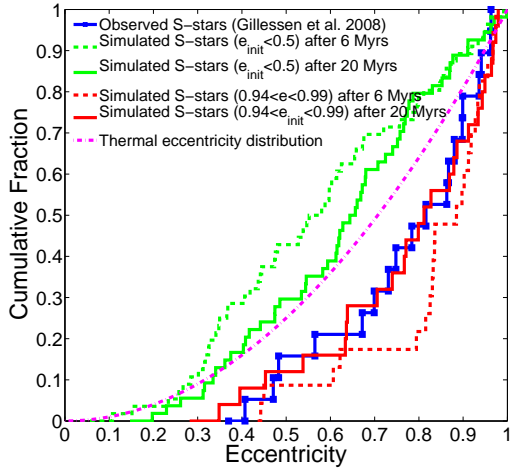} 

\caption{{\footnotesize Cumulative distribution of observed and simulated S-stars
eccentricities, for various models (see legend).}}

\label{f:eccentricities-cum} 
\end{figure}

In Figure~\ref{f:eccentricities-cum} we show the final cumulative
eccentricity distribution of the S-stars for the different origin
models (Table \ref{t:models}). These are compared to the the orbits
of the observed S-stars (taken from \citealp{gil+08}). The probabilities
for the samples of the observed and simulated S-stars to be randomly
chosen from the same distribution (calculated using a two-sample $\chi^{2}$
test) are given in Table \ref{t:models}. We find that the binary
disruption model taking place at least $20$ Myr ago is much favored
over all other models tested here. We find $93\%$ chance for the
observed S-stars and the simulated stars in such model to originate
from the same distribution (with the shorter timescale of $6$ Myr
consistent at the $26\%$ level) . In contrast, the disk migration
scenarios seem to be excluded (for the given assumptions), since they
have major difficulties in explaining the large fraction of eccentric
orbits observed for the S-stars in the GC.

We find that the inclination distribution of the captured stars, initialized
with the same inclination, is rapidly isotropized, to resemble a random
distribution of inclinations (consistent with random at the $\sim65\%$
and $\sim20\%$ level, after evolution of $20$ and $6$ Myr, respectively).
This is expected from the RR process, as discussed in the previous
section (see also fig. \ref{f:RR}). We conclude that the observed
isotropic distribution of the S-stars angular momentum direction is
consistent with all the S-stars production models studied here, and
can not be used to discriminate between them, although it constrains
the lifetime of the S-stars system to be at least $\sim4$ Myr at
the $95\%$ level, assuming all S-stars were initially put on the
same plane.

\subsection{Survival of the S-stars: tidal disruption, ejection and hypervelocity
stars}

As already discussed above, the S-stars can change their orbits due
to their dynamical evolution. A star could therefore be scattered
very close to the MBH and be disrupted by it, if its pericenter distance
from the MBH becomes smaller than the tidal radius of the star $r_{t}=r_{\star}(M_{\bullet}/m_{\star})^{1/3}$.
Many of the S-stars could therefore not survive for long close to
the MBH. We followed the orbits of stars in our simulations and calculated
the fraction of stars that have been disrupted. For the tidal disruption
calculations all stars were assumed to have the typical main sequence
radius according to their mass. We consider a star as being disrupted
if its pericenter became smaller than twice the tidal radius during
the simulation (i.e., when it is strongly affected by the MBH tidal
forces or even totally disrupted in one pericenter passage). We find
that most of the S-stars survived to current times in all the models
(see survival fractions in \ref{t:models}). The S-stars population
in the GC therefore gives a good representation of all the S-stars
formed/captured in this region. The production rate of the S-stars
required to explain current observations is therefore only slightly
higher ($1.2-1.3$ times higher) than that deduced from current number
of S-stars observed.

In principle the S-stars could be ejected by strong encounters to
orbits with larger semi-major axes, putting them outside the 0.05
pc region near the MBH \citep{mir+00}, or even ejecting them as unbound
hypervelocity stars \citep{ole+07}. The softening radius used in
our simulation was $r_{soft}=4R_{\odot}$, comparable to the radius
of observed S-stars, allowing us to follow even very close encounters.
Nevertheless, we find that only $\sim10\%$ of the stars were ejected
outside of the central $0.05$ pc, and even those had maximal semi-major
axis in the end of the simulation not extending beyond 0.1 pc. We
therefore conclude that such ejected S-stars can not explain recent
observations of many B-type stars outside the central $0.05$ pc (H
Bartko, private communication). Moreover, none of the $3\, M_{\odot}$
stars in our simulations have been ejected as a hypervelocity star,
suggesting that ejection of hypervelocity stars through encounters
with SBHs is not an efficient mechanism (see also \citealp{per09}
for the related constraints on this mechanism). We note that since
our simulations can be rescaled, we can probe much higher stellar
densities and smaller softening radius (see next section). When such
rescaling is used, in which all the 1200 stars are distributed between
$3\times10^{-4}$ pc to $0.1$ pc (rescaling the radii by half), we
find $5$ stars ejected beyond $0.1$ pc (to have final semi-major
axis of $0.1,\,0.16,\,0.16,\,0.21$ and $0.37$ pc), but none ejected
as hypervelocity stars or even as slow unbound stars.

\section{Dependence of the results on the assumed density of SBHs}

As discussed above, the density of the SBHs that are responsible for
the evolution in our $N$-body models is not well determined. Scaling
the $N$-body results to different assumed values of $N$ is complicated
by the fact that scalar resonant relaxation has two regimes, coherent
and random walk. In our models, the transition occurs at \begin{equation}
t_{M}\simeq QP/N\sim3\times10^{4}\left(\frac{N_{\mathrm{SBH}}}{10^{3}}\right)^{-1}\left(\frac{P}{100\,\mathrm{{\rm \textrm{yr}}}}\right)\,{\rm \textrm{yr}\,},\end{equation}
where the $N_{\mathrm{SBH}}^{-1}$ scaling is for scattering by SBHs
of a given mass.

In the coherent regime, $\Delta t\lesssim t_{M}$, orbital angular
momenta grow as

\begin{equation}
\frac{\Delta J}{J_{c}}\sim10^{-4}\left(\frac{N_{{\rm SBH}}}{10^{3}}\right)^{1/2}\frac{\Delta t}{P}\,.\end{equation}
 In the diffusive regime, $\Delta t\gtrsim t_{M}$, \begin{equation}
\left|\frac{\Delta J}{J_{c}}\right|\sim\sqrt{\frac{\tau}{\tau_{sRR}}}\approx2\times10^{-3}\sqrt{\frac{\Delta t}{P}}\,,\end{equation}
 independent of $N_{\mathrm{SBH}}$.

We are interested in the orbital evolution of the S-stars over timescales
of $\Delta t\sim O(10^{7})$ yr. In our simulations and those with
larger $N$, $\Delta t\gg t_{M}$. In this large-$N$ regime, changes
in eccentricity are dominated by the diffusive relation and are therefore
expected to be nearly independent of $N$ for time scales of interest,
at least up to $N$-values of $\sim10^{5}$ where the distributed
mass begins to approach the mass of the MBH and resonant relaxation
is no longer effective. Only for $N_{\mathrm{SBH}}\lesssim10$ does
$t_{M}$ approach $10^{7}$ yr and our results start depending significantly
on $N_{\mathrm{SBH}}$. However, such a small number of SBHs in the
volume of interest is highly unlikely, and therefore our results are
robust to the details of the SBH cusp model.

The $N$-body simulations can also trivially be rescaled by \begin{equation}
r\rightarrow Ar,\ \ \ \ t\rightarrow A^{3/2}t\end{equation}
 at fixed mass. This corresponds to placing the same number of SBHs
into a smaller (larger) region and integrating for a shorter (longer)
time. For instance, if we rescale our simulations to three times smaller
distances (in which case the stars are distributed between $3\times10^{-4}$
and $0.05$ pc), the integration time becomes $\sim5$ Myr.

\section{Summary}

The approximately thermal eccentricity distribution of the S-stars
near the massive black hole (MBH) in the galactic center (GC), $N(<e)\propto e^{2}$,
is not naturally predicted by either of the two leading models for
their production: migration of stars formed in a gaseous disk; or
capture of stars following binary disruption by the MBH. The former
model predicts eccentricities that are too low, the latter too high.
In this paper, we followed the dynamical evolution of orbits of various
eccentricities near the GC MBH, including for the first time the cluster
of stellar-mass black holes (SBHs) that is expected to form around
the MBH via mass segregation. We found that perturbations from the
SBHs can reduce the eccentricities of initially very eccentric orbits
($0.94\le e_{init}\le0.99$) into a distribution that is consistent
with the observed one, in a time of approximately $20$ Myr, comparable
to S-star lifespans; some of the stars change their eccentricities
by more than $0.5$ to values as low as $e_{final}=0.2-0.4$. We confirmed
these $N$-body results via a theoretical analysis of the relaxation
process, and used that analysis to argue that our results are not
strongly dependent on the (unknown) normalization of the SBH density
near the MBH. The same mechanism is unable to convert initially low-eccentricity
orbits into very eccentric ones on the same time scale, arguing against
the validity of the disk migration model for the origin of the S-stars.
We also found that most S-stars are not disrupted by the MBH during
their lifetime, and very few are ejected outside the central $0.05$
pc near the MBH, and none having a semi-major axis beyond $0.1$ pc.
We did not find any hypervelocity star ejected in our simulation.

Evolution toward a thermal eccentricity distribution is a natural
consequence of random gravitational encounters with a population of
massive perturbers. In this paper, we considered the effect of a background
of SBHs, which are expected on the basis of very general arguments
to contribute a total mass of $\sim10^{4}M_{\odot}$ in the inner
0.1 pc around the MBH. We showed that they were effective at moderating
the eccentricities of initially highly eccentric orbits. These results
strengthen models in which the S-stars form from disrupted binaries,
and disfavor models in which the S-stars are formed with low eccentricities.

\acknowledgements{}

This work was supported by grants NASA NNX07AH15G, NSF AST-0821141,
and NSF AST-0807810 to D. M., and by ISF grant 928/06 and Starting
Grant 202996 to T. A.

\bibliographystyle{apj}

\end{document}